\documentstyle[prb,aps,epsfig,multicol]{revtex}

\begin{document}
\draft
\preprint{HD-THEP-01-3}
\title{Equation of State for Helium--4 from Microphysics}
\author{Tim Gollisch$^{1,2}$ and Christof Wetterich$^{1}$}
\address{
$^1$Institut f\"ur Theoretische Physik,
Universit\"at Heidelberg,
Philosophenweg 16, 69120 Heidelberg,
Germany
\\
$^2$Innovationskolleg Theoretische Biologie,
Humboldt--Universit{\"a}t zu Berlin,
Invalidenstr.\ 43, 10115 Berlin,
Germany
}
\date{April 8, 2002}
\maketitle
\begin{abstract}
We compute the free energy of helium--4 near the lambda transition based on an exact renormalization--group equation.  An approximate solution permits the determination of universal and nonuniversal thermodynamic properties starting from the microphysics of the two--particle interactions.  The method does not suffer from infrared divergences.  The critical chemical potential agrees with experiment.  This supports a specific formulation of the functional integral that we have proposed recently.  Our results for the equation of state reproduce the observed qualitative behavior.  Despite certain quantitative shortcomings of our approximation, this demonstrates that \emph{ab initio} calculations for collective phenomena become possible by modern renormalization--group methods.
\end{abstract}

\begin{multicols}{2}

\section{Introduction}

When we regard a multiparticle system such as helium--4, we can choose between two possible points of view.  We can zoom in on the details of the interactions, focusing on only a few particles, and in a first approach we might choose to study how the interactions of just two particles can be described.  This is typically investigated by scattering experiments, and a quite precise picture of the two--particle interaction for helium--4 exists.\cite{Aziz}  On the other hand, we can also take a broader view and turn to the thermodynamic properties of many particles where properties such as pressure, temperature, and superfluidity arise.\cite{London,Donnelly,Wilks}

These two pictures are, of course, connected to each other as the thermodynamic properties are large scale effects of the (local) microscopic interactions.  Yet for helium--4 in the interesting region around the superfluid phase transition, a complete mathematical description, which makes this connection explicit, is still lacking, to our best knowledge.  A general problem in the calculation of thermodynamic properties is the occurrance of infrared divergences.  Here, we present a novel approach, which is free of those divergences.  It is based upon an exact renormalization--group equation.\cite{Wetterich}

The goal of this work is threefold.  First of all, we demonstrate how nonuniversal properties of thermodynamic systems can be calculated by means of an exact renormalization--group equation.  Second, we present first results for the critical chemical potential and the equation of state for helium--4 obtained merely from the microscopic interactions.  And third, we show the necessity of a shift of the chemical potential in the action, which arises from the mathematical manipulations of the functional integral and becomes important in the calculation of the phonon spectrum and the critical chemical potential.  We have reported on the appearance of this shift earlier \cite{GollischWetterich} and now show its effect in application.

The renormalization group is well suited for the treatment of the superfluid phase transition.  While a perturbative approach is plagued by infrared divergences, the renormalization group naturally circumvents these difficulties.  It also presents an intuitive connection between the microscopic and the macroscopic pictures.  It starts with just the microscopic interactions.  After integrating out the single--particle degrees of freedom consecutively from the effective ultraviolet (UV) cutoff at large momentum down to zero momentum, a complete thermodynamic description of the system is obtained.

This basic idea\cite{Wilson} is implemented in our approach by a functional differential equation that represents the lowering of an infrared cutoff.  There is, in principle, no restriction on the relative interaction strength or the density of the system as in most perturbative approaches.  Details of the method can be found elsewhere.\cite{Review}

A similar approach to the calculation of nonuniversal properties for Bose condensates by renormalization--group methods was taken by Bijlsma and Stoof, restricted to low density gases.\cite{BijlsmaStoof}  Another recent application of renormalization group methods to superfluid helium combined with the exploitation of the underlying symmetry and related Ward identity has successfully treated the infrared divergences that occur in the computation of low--lying excitations.\cite{Castellani}

\section{The action for helium--4}

We describe helium--4 by the many--body bosonic Hamiltonian operator
\begin{equation}
  \label{eq:H_op_ft}
  {\cal H}=\sum_q\left(\frac{q^2}{2m}-\mu\right)a^\dag_q a_q+\frac{1}{2}\!\sum_{q_1,q_2,q}\!a^\dag_{q_1+q}a^\dag_{q_2-q}v(q)a_{q_2}a_{q_1}.
\end{equation}
The mass $m$ of the helium atom is $3.73$~GeV, $\mu$ is the chemical potential, and $v(q)$ the two--particle interaction potential.

Such a description is clearly not valid below the atomic scale, and we therefore assume a UV momentum cutoff proportional to the inverse atomic length scale given by the atomic diameter $\sigma$:  $\Lambda=(2\pi)/\sigma$, $q^2<\Lambda^2$.  (In all formulas, units are chosen such that $\hbar=c=k_B=1$.)  We choose to describe the interaction by a simple Lennard--Jones potential
\begin{equation}
v(r)=4\epsilon\left(\frac{\sigma^{12}}{r^{12}}-\frac{\sigma^6}{r^6}\right)
\end{equation}
with the maximum energy of attraction $\epsilon=10.22$~K and $\sigma=2.556\mbox{ \AA}$.  This fits the measured interaction potential very accurately.\cite{Aziz}  For the definition of the interaction in the momentum domain, we identify the atomic diameter $\sigma$ with the scattering length\cite{Olinto} and cut off the potential at small distances while fixing $v(q\!=\!0)=(4\pi \sigma)/(Vm)$ as required by low energy scattering theory.  Here, $V$ is the system's volume, which we will let eventually approach infinity.  The Fourier transform of the two--particle potential thus derived is shown in Fig.~\ref{fig:potential}.
\noindent
\parbox{\linewidth}{
\begin{figure}
\epsfig{file=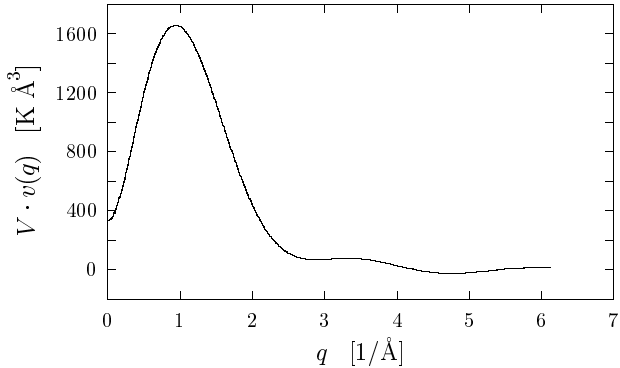, width = \linewidth}
\caption{Fourier transform of the two--particle interaction for helium--4.
\label{fig:potential}}
\end{figure}
}

The action ${\cal S}[\chi,\chi^*]$ for helium--4, which is a functional depending on the four--dimensional classical complex field $\chi$, is defined via the grand canonical partition function
\begin{equation} \label{eq:partitionFunction}
\mbox{Tr}\;e^{-\beta {\cal H}}=\int {\cal D} \chi {\cal D} \chi^* e^{-{\cal S}[\chi,\chi^*]}
\end{equation}
with $\beta=1/T$ denoting the inverse temperature.  As we have shown earlier,\cite{GollischWetterich} great care has to be taken in constructing the action, and for the Hamiltonian operator (\ref{eq:H_op_ft}), we have obtained a perhaps unexpected shift $\Omega$ for the chemical potential $\mu$ in the action.  One finds (up to a constant)
\begin{eqnarray} \nonumber
\lefteqn{{\cal S}[\chi,\chi^*]} \\ \nonumber
&=&\int_0^\beta d \tau \bigg\{\sum_q\chi_q^*(\tau)\left(\frac{\partial}{\partial\tau}+\frac{q^2}{2m}-\mu-\Omega\right)\chi_q(\tau) \\ \label{eq:action}
  & &+\;\frac{1}{2}\!\!\sum_{q_1,q_2,q} \!\!\!
      \chi^*_{q_1+q}(\tau)\chi^*_{q_2-q}(\tau)v(q)\chi_{q_2}(\tau)\chi_{q_1}(\tau)
    \!\bigg\}
\end{eqnarray}
with
\begin{equation}
\Omega=\frac{1}{2}\sum_q\left[v(q)+v(0)\right].
\end{equation}
All momenta are restricted by the UV cutoff $q^2 < \Lambda^2$.

It is convenient to express the field as a sum over Matsubara frequencies $\omega_n=2\pi n/\beta$
\begin{equation}
\chi_q(\tau)=\sum_{n=-\infty}^\infty e^{i\omega_n\tau} \chi_{n,q},
\end{equation}
such that
\begin{eqnarray} \nonumber
{\cal S}[\chi,\chi^*]&=&\sum_{n,q}\beta\chi^*_{n,q}\left(i\omega_n+\frac{q^2}{2m}-\mu-\Omega\right)\chi_{n,q} \\ \nonumber
&&+\frac{\beta}{2}\sum_{n_1,\ldots,n_4 \atop q_1,q_2,q} \chi^*_{n_1,q_1+q}\chi^*_{n_2,q_2-q} \\ \label{eq:classicalAction}
&& \;\;\;\;\;\;\;\;\;\;\;\times \chi_{n_3,q_2}\chi_{n_4,q_1} v(q)\delta_{n_1+n_2,n_3+n_4}.
\end{eqnarray}

This action differs in the appearance of $\Omega$ from a functional--integral approach proposed earlier within a coherent--state ansatz, where a complete set of eigenstates of the annihilation operators is used for the conversion of the partition function to a functional integral.\cite{Casher}  In contrast, Eq.~(\ref{eq:classicalAction}) is obtained by transforming the creation and annihilation operators in the Hamiltonian (\ref{eq:H_op_ft}) to location and momentum operators before the conversion to a functional integral.  This allows us to choose an operator ordering that avoids uncontrollable contributions from next--to--leading--order terms in the discretization parameter.  The correctness of our approach has been tested by numerical calculations in quantum mechanics.\cite{GollischWetterich}

\section{Phonon spectrum}

\label{sec:phononSpectrum}

From the action (\ref{eq:classicalAction}), we can already calculate the phonon velocity for a weakly coupled Bose condensate at $T=0$.  Altough this can also be obtained by a Bogoljubov transformation of the Hamiltonian operator (\ref{eq:H_op_ft}), we sketch the calculation as it shows how $\Omega$ cancels a corresponding term evolving from the one--loop correction.  In this way, $\Omega$ acts as a counterterm though it has naturally emerged from the construction of the functional integral.

The generating functional of the connected Green functions $W[J,J^*]$ is defined by
\begin{eqnarray} \nonumber
  e^{W[J,J^*]} &=&\int{\cal D}\chi{\cal D}\chi^* \exp\Bigg\{-{\cal S}[\chi,\chi^*] \\ \label{eq:expW}
&&+\sum_{n,q}\left(J_{n,q}^*\chi_{n,q}+J_{n,q}\chi_{n,q}^*\right)\Bigg\},
\end{eqnarray}
and the generating functional of the 1\emph{PI} Green functions by the Legendre transform
\begin{equation}
  \Gamma[\varphi,\varphi^*]=-\ln W[J,J^*]+\sum_{n,q}\left(J_{n,q}^*\varphi_{n,q}+J_{n,q}\varphi_{n,q}^*\right).
\end{equation}
To calculate the phonon spectrum, we need the zeros of the inverse full propagator $\Gamma^{(2)}$.  We restrict ourselves to the case of a static and homogeneous condensate $\overline{\varphi}_{n,q}=\varphi\,\delta_{n,0}\,\delta_{q,0}$.  Considering only low--momentum excitations, we can approximate the interaction potential $v(q)$ by $v(0)$.  The tree--level approximation yields
\begin{equation}
\Gamma_0[\varphi,\varphi^*]={\cal S}[\varphi,\varphi^*]
\end{equation}
and therefore (for fixed $n$ and $q$)
\begin{eqnarray} \nonumber
\Gamma_{0}^{(2)}
&=&
\left(
\begin{array}{cc}
\frac{\delta^{2}{\cal S}[\varphi,\varphi^{*}]}{\delta\varphi^{*}_{n,q} \delta\varphi_{n,q}} &
\frac{\delta^{2}{\cal S}[\varphi,\varphi^{*}]}{\delta\varphi_{-n,-q} \delta\varphi_{n,q}} \\
& \\
\frac{\delta^{2}{\cal S}[\varphi,\varphi^{*}]}{\delta\varphi^{*}_{n,q} \delta\varphi^{*}_{-n,-q}} &
\frac{\delta^{2}{\cal S}[\varphi,\varphi^{*}]}{\delta\varphi_{-n,-q} \delta\varphi^{*}_{-n,-q}} \\
\end{array}
\right) \\ \label{eq:gamma_0^2}
&=&
\beta \left(
\begin{array}{cc}
i\omega_{n} + \tilde{q} & v(0){\varphi^{*}}^{2} \\
& \\
v(0)\varphi^{2} & -i\omega_{n} + \tilde{q} \\
\end{array}
\right),
\end{eqnarray}
where
\begin{equation}
\tilde{q}=q^2/(2m)-\mu-\Omega+2v(0)|\varphi|^2.
\end{equation}
The one--loop correction is obtained as
\begin{equation}
\Gamma_1[\varphi,\varphi^*]=\frac{1}{2}\mbox{Tr}\; \ln \Gamma_0^{(2)}
\end{equation}
with the trace including a sum of $n$ and momenta.
Neglecting terms of ${\cal O}\left(v(0)^2\right)$, we obtain in the limit $\beta\rightarrow\infty$ the simple expression for the leading term,
\begin{equation}
\Gamma_1^{(2)}=\left(
\begin{array}{cc}
\beta\Omega & 0 \\
0 & \beta\Omega \\
\end{array}
\right).
\end{equation}
This cancels the corresponding term in $\tilde{q}$ appearing in Eq.~(\ref{eq:gamma_0^2}).

For $\Gamma_0^{(2)}+\Gamma_1^{(2)}$, we can now compute the eigenvalues:
\begin{equation}
  \lambda_{1,2}
=
\beta\Bigg(\frac{q^{2}}{2m}-\mu+2v(0)\left|\varphi\right|^{2}
\pm\sqrt{(i\omega_{n})^{2}+v(0)^{2}|\varphi|^{4}\;}\Bigg).
\end{equation}
The phonon energy spectrum is determined by the zeros of $\lambda$ for complex $\omega_n$ (with the first--order approximation\cite{LandauLifshitz} $\mu=v(0)\left|\varphi\right|^2$):
\begin{equation}
E(q)=i\omega_n(q)=\left|q\right|\;\;\cdot\sqrt{\frac{1}{m}v(0)|\varphi|^2\;} + {\cal O}(q^2)
\end{equation}
with the correct phonon velocity of a weakly coupled Bose condensate at $T=0$.\cite{LandauLifshitz}

\section{Flow equation}

The physical situation of vanishing sources corresponds to an extremum of $\Gamma$ since $\delta \Gamma / \delta \varphi = J^*$.  The value of $\Gamma$ at the extremum $\Gamma_{\rm eq}=\Gamma[\varphi_{\rm eq},\varphi^*_{\rm eq}]$ is directly related to the free energy $F$ by
\begin{equation}
F=T\,\Gamma_{\rm eq}+\mu\, N.
\end{equation}
For a computation of thermodynamic quantities, we therefore aim at a computation of $\Gamma_{\rm eq}$ as a function of $T$ and $\mu$.  This involves a complicated functional integral over fluctuations (\ref{eq:expW}).  We proceed by a stepwise solution by introducing a cutoff $k$ for the fluctuations such that only fluctuations with momenta $q^2 > k^2$ are included in the functional integral.  The variation of the effective action with $k$ is described by an exact flow equation.

For notational simplicity, we rescale the fields by absorbing a factor of $\sqrt{\beta/(2mV)}$ into the field variable and define $\tilde{\mu}=2m(\mu+\Omega)$ and $\tilde{v}(q)=4m^2V v(q)/\beta$.  For the action this yields
\begin{eqnarray} \nonumber
  {\cal S}[\chi,\chi^*]
&=&V\sum_{n,q}\chi_{n,q}^*\left(2im\omega_n+q^2-\tilde{\mu}\right)\chi_{n,q} \\ \nonumber
  & &+\frac{V}{2}\!\!\sum_{n_1,\ldots,n_4 \atop q_1,q_2,q}\chi_{n_1,q_1+q}^*\chi_{n_2,q_2-q}^* \\ \label{eq:classicalActionTransformed}
&& \;\;\;\;\;\;\;\;\;\;\;\times\chi_{n_3,q_2}\chi_{n_4,q_1}
  \tilde{v}(q)\delta_{n_1+n_2,n_3+n_4}.
\end{eqnarray}

In the grand canonical partition function, we add source terms $J$ and an infrared regulator $R_k(n,q)$ and define the functional $W_k[J,J^*]$ as the generating functional of the connected Green functions in the presence of an infrared cutoff $k$,
\begin{eqnarray} \nonumber
e^{{W}_k[J,J^*]}
&=&
\int{\cal D}\chi{\cal D}\chi^*\exp\Bigg\{-{\cal S}[\chi,\chi^*]
\\ \nonumber
&&+\sum_{n,q}\Big(J_{n,q}^*\chi_{n,q}
+J_{n,q}\chi_{n,q}^* 
\\ \label{eq:funcIntCutoff}
&& \;\;\;\;\;\;\; -V\chi_{n,q}^*R_k(n,q)\chi_{n,q}\Big)\!\Bigg\}.
\end{eqnarray}
For the infrared regulator, we choose
\begin{equation}
  \label{eq:R_k(q)}
  R_k(n,q)=\frac{Z_kq^2}{\exp\left(\frac{q^2}{k^2}\right)-1}\;\delta_{n,0}.
\end{equation}
Here, $Z_k$ is a wave--function renormalization, which will be defined below in Eq.~(\ref{eq:truncation}).  We note that this particular regulator cuts off the momentum modes with $q^2<k^2$ only for the $n=0$ Matsubara frequency.  The $n\ne 0$ modes are not affected and contribute fully in the functional integral (\ref{eq:funcIntCutoff}).  For large enough $T$, the integration of the $n\ne 0$ frequencies poses no problem since temperature acts as an infrared regulator for these modes.  Nevertheless, a future extension of the infrared cutoff to the $n\ne 0$ Matsubara modes may be welcome for low $T$.

The effective average action is defined as
\begin{eqnarray} \nonumber
\Gamma_k[\varphi,\varphi^*]
&=&-{W}_k+\sum_{n,q}\Big[J_{n,q}^*\varphi_{n,q}+J_{n,q}\varphi^*_{n,q} \\ 
&& \;\;\;\;\;\;\;\;\;\;\;\;\;\; -V\varphi^*_{n,q}R_k(n,q)\varphi_{n,q}\Big].
\end{eqnarray}
As the infrared cutoff $k$ is lowered, the evolution of $\Gamma_k$ follows the exact renormalization group equation\cite{Review}
\begin{equation}
  \label{eq:ERGE}
    \partial_{k} \Gamma_{k} \left[\varphi^{*},\varphi \right] \bigg|_{\varphi^{*},\varphi} =
\frac{1}{2} \mbox{Tr} \left( \left( \partial_{k} R_{k} \right) \left( \Gamma_{k}^{(2)} + R_{k} \right)^{-1} \right).
\end{equation}

The functional differential equation (\ref{eq:ERGE}) can only be solved approximately by truncating the most general functional form of the effective average action.  For a truncation scheme, we choose a derivative expansion
\begin{eqnarray} \nonumber
\Gamma_k[\varphi^*,\varphi] &=&
\int_0^\beta \frac{d\tau}{\beta} \int d^3x \Bigg[ U_k(\rho)
 \\ \nonumber
&& + \varphi^*(\tau,x)\left(2m\frac{\partial}{\partial \tau}-Z_k \nabla^2\right)\varphi(\tau,x) \\ \label{eq:truncation}
&& + \frac{1}{4} Y_k(\nabla \rho)(\nabla \rho)\Bigg].
\end{eqnarray}
We consider an arbitrary dependence of the effective potential $U_k(\rho)$ on the invariant $\rho(\tau,x)=\varphi^*(\tau,x)\varphi(\tau,x)$.  The derivative terms (``kinetic terms'') are multiplied by wave--function renormalizations $Z_k$ and $Y_k$.  The momentum and field dependence of $Z_k$ and $Y_k$ are expected to be weak \cite{Review} and therefore neglected for the computation of ground--state properties at this stage.  For the calculation of the excitation spectrum, these should turn out to be important and thus should be included in future calculations.

We obtain evolution equations for the effective potential and its derivatives by expanding the renormalization--group equation (\ref{eq:ERGE}) around a constant background field and taking the appropriate derivatives with respect to $\rho$.  In particular, we will use the evolution equation for the first derivative of the effective potential $U_k^\prime=\partial U_k / \partial \rho$, which is obtained as
\begin{eqnarray} \nonumber
\lefteqn{\partial_t U^\prime_k(\rho)} \\ \nonumber
& = &
-\frac{1}{2} \int \frac{d^3q}{(2\pi)^3} \Big( \partial_t R_k(q) \Big) \Bigg(
\frac{U_k^{\prime\prime}}{\left(Z_k q^2 + U_k^{\prime} + R_k(q)\right)^2} \\ \label{eq:dUprimedt}
&& +\frac{3U_k^{\prime\prime}+2U_k^{\prime\prime\prime}\rho+Y_k q^2}{\left(Z_k q^2 +Y_k\rho q^2+ U_k^{\prime} + 2 U_k^{\prime \prime} \rho + R_k(q)\right)^2} \Bigg).
\end{eqnarray}
Evolution equations for $Z_k$ and $Y_k$ are obtained by taking further appropriate expansions of Eq.~(\ref{eq:ERGE}).\cite{Review,TetradisWetterich}

The solution of $U_k$ for $k\rightarrow 0$ is directly related to intensive thermodynamic quantities as pressure $P$, energy density $\epsilon$, or particle density $n$.  Denoting by $U_{\rm eq}(T,\mu)$ the value of $U_{k=0}$ at its minimum, one finds the relations
\begin{equation}
P=-TU_{\rm eq},\;\;\;n=-T\frac{\partial U_{\rm eq}}{\partial \mu},\;\;\;
\epsilon=-T^2\frac{\partial U_{\rm eq}}{\partial T}+\mu n.
\end{equation}
Therefore $U_{\rm eq}(T,\mu)$ contains the information about the equation of state.  Furthermore, the correlation length is encoded in the derivatives of $U_{k=0}(\rho)$ at the minimum.  We therefore aim for a solution of the flow equation (\ref{eq:dUprimedt}) (or a similar equation for $U_k$) for $k\rightarrow 0$.

This is obtained numerically by discretizing $\rho$ and using discrete approximations for the $\rho$ derivatives of $U_k$.  As $U_k(\rho)$ remains fairly smooth throughout the evolution, a relatively small number of sampling points suffice to give a good approximation of the shape of $U_k$.  The results presented in Sec.~\ref{sec:results} were obtained with 20 sampling points in $\rho$ space.  In contrast to a Taylor expansion of $U_k$ around its minimum, this retains more information about the influence of regions of the effective potential away from the minimum.  The resulting system of coupled ordinary differential equations was solved by an adaptive--stepsize Runge--Kutta integrator, where the momentum integrals were also evaluated numerically at each step.

The generic evolution of $U_k(\rho)$ is discussed in detail elsewhere.\cite{Review}  In short, one observes that the minimum of $U_k(\rho)$ moves towards zero, but converges to a finite value $\rho_0$ for the superfluid phase with $U_0^\prime(\rho_0) = 0$.  Furthermore, $U_k(\rho)$ is found to level out in the region $\rho<\rho_0$ in accordance with the requirement that the true effective potential at $k=0$ must be a convex function of $|\varphi|$.  For the nonsuperfluid phase, on the other hand, the minimum of $U_0$ is at $\rho_0=0$ with $U_0^\prime > 0$.

The finite number of sampling points is a possible source of numerical inaccuracies, especially in the approach of the nonanalyticity at $\rho_0$ in the superfluid phase and at the boundary $\rho=0$ in the nonsuperfluid phase.  But since changing the number of sampling points does not substantially affect the results and since the critical exponents are calculated with high accuracy as seen later, we do not expect these numerical sources to be major contributions to the errors of the results.

The critical line separating the superfluid and non--superfluid phase corresponds to a ``fixed--point'' or ``scaling'' solution of the evolution equation in the dimensionless variables $\tilde{\rho}=Z_k k^{-1} \rho$ and $u_k(\tilde{\rho})=k^{-3} U_k(\tilde{\rho})$.  It is given by $\partial_k u_k(\tilde{\rho})|_{\tilde{\rho}}=0$.
Figure~\ref{fig:flow} shows an example of the flow of the location $\kappa$ of the minimum of $u_k(\tilde{\rho})$ for two different temperatures above and below the critical temperature.
\noindent
\parbox{\linewidth}{
\begin{figure}
\epsfig{file=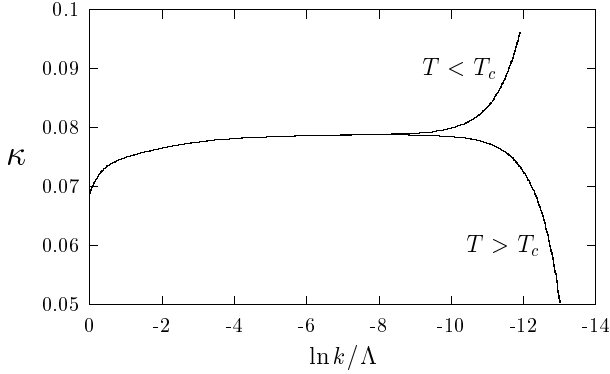, width = \linewidth}
\caption{Flow of the location of the minimum of the dimensionless effective potential $u_k(\tilde{\rho})$ close to the critical temperature ($T_c=2.172$~K, $\left| T-T_c \right| \approx 10^{-7}$~K).  One observes the approximate fixed point in the range $\ln (k/\Lambda)\approx -4$ to $-10$.
\label{fig:flow}}
\end{figure}
}
We see that $\kappa$ stays close to the fixed--point solution for several orders of magnitude of $k$.  Above the critical temperature, it finally runs towards the symmetric phase where the minimum of the effective potential is at zero.  Below the critical temperature, the minimum eventually runs off in the other direction leading to a finite expectation value of the field and therefore to the case of spontaneously broken symmetry.

\section{Dimensional reduction}

Since our cutoff acts only on the $n=0$ Matsubara frequency, the ``initial value'' of $\Gamma_k$ at $k=\Lambda$ has to be computed as an integral over the $n\ne 0$ frequencies.  This procedure is known as \emph{dimensional reduction}.  It adds to the classical potential
\begin{equation}
\label{eq:startU}
U_\Lambda^{(0)}(\rho)=-\tilde{\mu}\rho+\frac{1}{2}\tilde{v}(0)\rho^2
\end{equation}
a correction.  The crudest approach would compute this in a one--loop approximation.  In order to get a little more accurate, we choose, instead, a description in terms of a standard Schwinger--Dyson equation.  This takes the leading mass renormalization into account.  The Schwinger--Dyson equation is obtained by a one--loop calculation of the inverse full propagator of the Goldstone mode, with renormalized vertices given by Eq.~(\ref{eq:truncation}).  It yields a self--consistent equation, which we solve numerically at discretized points in $\rho$ space for $U^\prime(\rho)$ while keeping the wave--function renormalizations and $U^{\prime\prime}(\rho)$ fixed at their original values.  This procedure is justified \emph{a posteriori} by the observation that the effect of the Schwinger--Dyson equation is nearly a constant shift of $U^\prime(\rho)$ for all $\rho$.

The Schwinger--Dyson equation is obtained in the following way.  We decompose the field $\varphi$ into its real and imaginary parts $\varphi=(\pi+i\sigma)/\sqrt{2}$ and write the truncation (\ref{eq:truncation}) as
\begin{eqnarray} \nonumber
  \Gamma[\pi,\sigma] &=& \int_0^\beta \frac{d\tau}{\beta}\int d^3x\Bigg\{U\left(\rho\right)+2im\cdot\pi\frac{\partial}{\partial\tau}\sigma \\
&& +\frac{1}{2} Z\left[(\nabla\pi)^2+(\nabla\sigma)^2\right] + \frac{1}{4} Y(\nabla\rho)^2\Bigg\}.
\end{eqnarray}
The inverse full propagator of the Goldstone mode is then given by
\begin{equation}
  \label{eq:gamma2_trunk}
  \left. \frac{1}{V}\frac{\delta \Gamma_\Lambda[\pi,\sigma]}{\delta \pi_{n,q} \delta \pi_{-n,-q}} \right|_{\pi = 0}
  =Z_\Lambda q^2+U_\Lambda^\prime(\rho).
\end{equation}
In order to calculate this propagator from the action (\ref{eq:classicalActionTransformed}), we use the identity
\begin{eqnarray} \nonumber
0 &=& \int {\cal D}\pi\, {\cal D}\sigma\; \frac{\delta}{\delta \pi_{n,q}} \exp \Bigg\{ - {\cal S}\left[ \pi, \sigma \right] \\
&& + \sum_{n,q} \left( J^{(\pi)}_{-n,-q}\cdot \pi_{n,q}
+ J^{(\sigma)}_{-n,-q} \cdot \sigma_{n,q} \right) \Bigg\},
\end{eqnarray}
where $J^{(\pi)}$ and $J^{(\sigma)}$ are source terms for $\pi$ and $\sigma$.  This leads us to
\begin{equation} \label{eq:calcOfJ}
0 = \left( -\frac{\delta{\cal S}}{\delta \pi_{n,q}}\left[ \frac{\partial}{\partial J^{(\pi)}}, \frac{\partial}{\partial J^{(\sigma)}} \right] + J^{(\pi)}_{-n,-q} \right) e^{W[J^{(\pi)},J^{(\sigma)}]}.
\end{equation}
We can now substitute the derivatives of $W$ by the renormalized vertices that we obtain from the truncation (\ref{eq:truncation}), e.g., by using the fact that $\Gamma^{(2)}$ is the inverse of the full propagator $W^{(2)}$.  We denote the propagators of the Goldstone and the radial modes by $P_\pi$ and $P_\sigma$:
\begin{eqnarray} \nonumber
P_\pi&=&Z_k q^2 + U_k^\prime(\rho) + R_k(n,q), \\ \label{eq:props}
P_\sigma&=&Z_k q^2 + Y_k\rho q^2 + U_k^\prime(\rho) +\! 2\rho U^{\prime\prime}(\rho) +\! R_k(n,q).\!
\end{eqnarray}
To account for the $n$ dependence of the propagators, we distinguish the propagators of the zeroth Matsubara frequency (with $R_k(n,q)\ne 0$) by writing $\hat{P}_\pi$ and $\hat{P}_\sigma$.

To obtain the propagator of the Goldstone mode, we furthermore need to take the derivative of the resulting expression for $J^{(\pi)}_{-n,-q}$ with respect to $\pi_{-n,-q}$.  We neglect two--loop graphs and explicitly evaluate the sums over Matsubara frequencies stemming from the $\delta S/\delta \pi$~term in Eq.~(\ref{eq:calcOfJ}).  This yields (for vanishing momentum and Matsubara frequency zero)
\begin{eqnarray} \nonumber
\lefteqn{\frac{1}{V}\frac{\delta^2 \Gamma_\Lambda}{\delta\pi_{n,q}\delta\pi_{-n,-q}}\Bigg|_{\pi=0,q=0,n=0}} \\ \nonumber
& = &
\frac{1}{V} \frac{\delta J_{-n,-q}}{\delta \pi_{-n,-q}} \Bigg|_{\pi=0,q=0,n=0} \\ \nonumber
  &=& -\tilde{\mu}+\tilde{v}(0)\rho
+\frac{1}{4\pi^2}\int_0^\Lambda dq\; q^2\left(\hat{P}_\pi\hat{P}_\sigma\right)^{-1} \\ \nonumber
&& \times \Bigg[\left[\tilde{v}(0)+2\tilde{v}(q)\right]\hat{P}_\sigma+\tilde{v}(0)\hat{P}_\pi
-4\rho U_\Lambda^{\prime\prime}(\rho)\tilde{v}(\rho) \Bigg] \\ \nonumber
&&+\frac{1}{4\pi^2}\int_0^\Lambda dq\; q^2\left(\frac{\frac{\beta}{4m}\coth \frac{\beta}{4m}\sqrt{P_\pi P_\sigma}}{\sqrt{P_\pi P_\sigma}}- \frac{1}{P_\pi P_\sigma}\right) \\ \label{eq:schwinger-dyson}
&& \times
\Bigg[\left[\tilde{v}(0)+2\tilde{v}(q)\right]P_\sigma+\tilde{v}(0)P_\pi
-4\rho U_\Lambda^{\prime\prime}(\rho)\tilde{v}(\rho)\Bigg] .
\end{eqnarray}
By equating the right--hand sides of Eqs.~(\ref{eq:gamma2_trunk}) and (\ref{eq:schwinger-dyson}), we obtain the Schwinger--Dyson equation, which we solve numerically for $U_\Lambda^\prime(\rho)$ with $U_\Lambda^{\prime\prime}(\rho)=\tilde{v}(0)$ on a discrete set of $\rho$ values.  This yields the initial condition for the evolution of $U_k^\prime(\rho)$.

A similar procedure could be used for a determination of $Z_\Lambda(\rho)$ and $Y_\Lambda(\rho)$ from the Schwinger--Dyson equation at nonzero $q^2$.  We omit here the fluctuation effects for the wave--function renormalization and use the ``classical'' values $Z_\Lambda=1$, $Y_\Lambda=0$.  We believe that this simplification gives a sizeable contribution to the error in the computation of nonuniversal quantities presented in Sec.~\ref{sec:results}.

We are aware that the use of the Schwinger--Dyson equation becomes problematic for low $T$.  In this region, one would prefer to use an infrared regulator that drops the Kronecker delta in Eq.~(\ref{eq:R_k(q)}).  As a result, also the contributions of the $n \ne 0$ Matsubara frequencies to the Schwinger--Dyson integral would be suppressed by the cutoff $\Lambda$, and all modes with $q^2 < \Lambda^2$ could then be dealt with by the renormalization--group procedure (in contrast to the present version where this holds only for $n=0$).  One would thus expect a more reliable treatment.  Modifying the cutoff~(\ref{eq:R_k(q)}) by dropping $\delta_{n,0}$ results again in a simple form of the evolution equation since the Matsubara sum can be evaluated explicitly.  We find
\begin{eqnarray} \nonumber
  \partial_k U_{k}(\rho)
&=&
\int \frac{d^3q}{(2\pi)^3} \left[\partial_k R_k(q)\right] \\ \label{eq:dUdt_unstable}
&& \times \frac{\beta}{8 m} \frac{P_\pi+P_\sigma}{\sqrt{P_\pi\cdot P_\sigma\;}}
\coth \frac{\beta}{4m}\sqrt{P_\pi\cdot P_\sigma\;}.
\end{eqnarray}
A numerical solution of this evolution equation, however, shows instabilities, which we have not managed to overcome so far.

\section{Results}

\label{sec:results}

From the solution of the evolution equation, we can calculate the quantities of interest for the phase transition to superfluidity.

{\bf Critical chemical potential.}
For a constant temperature of $2.172$~K, which corresponds to the experimental critical temperature under vapor--pressure conditions, we can tune the chemical potential to obtain the fixed--point solution.  We thereby find a critical chemical potential $\mu_c=-6.7$~K in good agreement with the experimental findings of Maynard,\cite{Maynard} where a value of around $-7.4$~K has been extrapolated from the measurements of the velocity of the fourth sound in the superfluid phase.  We emphasize that the $\Omega$ shift in Eq.~(\ref{eq:classicalAction}) is essential, as otherwise we would have obtained a value of about $12$~K higher.  This supports again our ansatz for the action for helium--4.

{\bf Critical exponents.}
By exploring the region around the critical temperature, we can compute the critical exponents $\nu$, $\beta$, and $\gamma$ of the helium--4 system.  They describe the divergence of the correlation length above and below the critical temperature ($\nu$), the growing of the order parameter, i.e., the expectation value of the field, below the critical temperature ($\beta$), and the divergence of the susceptibility above and below the critical temperature ($\gamma$).  We present our results as a demonstration that the flow equations can indeed be followed directly from microphysics to macrophysics without encountering infrared problems.  No resummations of series or other technical tricks as used by other analytical methods are needed.

It is well established that helium--4 at the lambda transition is described by the $O(2)$ universality class.  The critical exponents can therefore be compared to previous high--precision estimates for this universality class.  The excellent correspondence with other, more demanding calculations and experimental values shown in Table~\ref{tab:criticalExponents} demonstrates that our truncation is reliable for the universal quantities.  The scaling requirement that $\nu$ and $\gamma$, respectively, obtain the same values above and below the critical temperature is also well fulfilled.
\noindent
\parbox{\linewidth}{
\begin{table}
\caption{Critical exponents for the superfluid phase transition of helium--4 and comparison with values for the $O(2)$ universality class.  The slight differences in the values obtained for $\nu$ and $\gamma$ above and below the critical temperature reflect numerical inaccuracies.
\label{tab:criticalExponents}}
\begin{tabular}{cccc}
& Our & Other & Experimental \\
& results & approaches\cite{ZJ,Kondor} & results\cite{Lipa,Goldner,Swanson} \\ \tableline
& & 0.6695\tablenotemark[3] & 0.67095(13) \\
$\nu$ & \raisebox{1.5ex}[-1.5ex]{0.675\tablenotemark[1]} & 0.671\tablenotemark[4] & 0.6705(6) \\
& \raisebox{1.5ex}[-1.5ex]{0.667\tablenotemark[2]} & 0.672\tablenotemark[5] & 0.6708(4) \\
&&& \\
& & 0.3455\tablenotemark[3] & \\
\raisebox{1.5ex}[-1.5ex]{$\beta$} & \raisebox{1.5ex}[-1.5ex]{0.359\tablenotemark[2]} & 0.3485\tablenotemark[4] & \\
&&& \\
& & 1.316\tablenotemark[3] & \\
$\gamma$ & \raisebox{1.5ex}[-1.5ex]{1.305\tablenotemark[1]} & 1.315\tablenotemark[4] & \\
& \raisebox{1.5ex}[-1.5ex]{1.298\tablenotemark[2]} & 1.33\tablenotemark[5] & \\
\end{tabular}
\tablenotetext[1]{For $T>T_c$.}
\tablenotetext[2]{For $T<T_c$.}
\tablenotetext[3]{From summed perturbation series at six--loop order.}
\tablenotetext[4]{From a fifth--order $\epsilon$ expansion.}
\tablenotetext[5]{From lattice calculations.}
\end{table}
}

{\bf Pressure.}
Since $\delta\Gamma/\delta\varphi^*=J$, the minimum of the effective potential $U_{k=0}(\rho_0)$ corresponds to the physical case of vanishing external sources for the stationary helium system.  The pressure $P$ is given by the value of the effective potential at the minimum $\rho_0$, $P=-T U_{k=0}(\rho_0)$.
The minimum value $U_k(\rho_0(k))$ can be followed by an evolution equation, which is obtained from Eq.~(\ref{eq:ERGE}).  It is given by
\begin{eqnarray} \nonumber
\lefteqn{\partial_k \Big(U_k(\rho_0(k))\Big)} \\ \nonumber
&=&\frac{1}{2} \int\! \frac{d^3q}{(2\pi)^3} \Big[ \partial_k R_k(q) \Big] \Bigg( \frac{1}{Z_k q^2 + R_k(q)} \\
&&  +
\frac{1}{Z_k q^2 \! + \! Y_k \rho_0(k) q^2 \! + \! 2 U_k^{\prime \prime}(\rho_0(k)) \rho_0(k) \! + \! R_k(q)}\!\! \Bigg)\!\!
\end{eqnarray}
in the spontaneously broken regime ($\rho_0(k)\ne 0$) and by
\begin{equation}
\partial_k \left(U_k(0)\right)
= \int\! \frac{d^3q}{(2\pi)^3} 
\frac{\partial_k R_k(q)}{Z_k q^2 + U_k^\prime(0) + R_k(q)}
\end{equation}
in the symmetric regime ($\rho_0(k)=0$).  The starting value is the classical minimum $U_\Lambda(\rho_0(\Lambda))=-\tilde{\mu}^2/[2\tilde{v}(0)]$.

\noindent
\parbox{\linewidth}{
\begin{figure}
\epsfig{file=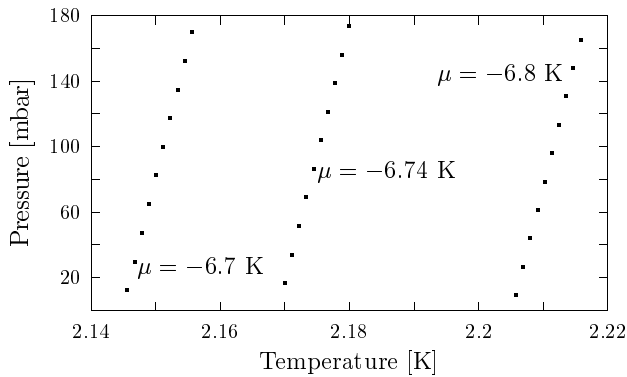, width = \linewidth}
\caption{Temperature dependence of the pressure for three different values of the chemical potential $\mu$ calculated from the evolution equation~(\ref{eq:dUprimedt}).  (The value of $\mu=-6.74$~K corresponds to the chemical potential at the critical temperature under vapor pressure as obtained from the fixed--point solution.)
\label{fig:pressure}}
\end{figure}
}
We find that the solution of the flow equations does not lead to a reliable estimate of the absolute value of the pressure.  The main reason for this is a strong dependence on the UV cutoff, which is not well controlled by our method.  For example, the initial value of $U_\Lambda(\rho_0(\Lambda))$ should also receive a correction from higher Matsubara frequencies, which we do not calculate in our Schwinger--Dyson approach.  This correction will depend notably on the microphysical details of the interaction and the choice of the UV cutoff.  Fortunately, this is not the case for the temperature dependence of the pressure, which is mainly governed by the momentum range $q^2\approx 2mT$.  We thus resort to normalizing the pressure at the critical temperature, where $\mu$ is given by the fixed--point condition, and only calculate the $T$ dependence of $P$.  Figure~\ref{fig:pressure} shows the dependence of the pressure on the temperature for three different values of the chemical potential.  Since we need to specify the chemical potential in our formalism when we want to calculate the order parameter and the density at vapor pressure, we determine $\mu(P,T)$ from this result and use the experimentally known vapor--pressure curve for $P(T)$.

{\bf Order parameter.}
While the critical exponent $\beta$ describes the growing of the order parameter in the $O(2)$ universality class, we can also calculate the order parameter for the helium system explicitly.  It is given by the expectation value of the field $\varphi$.  Therefore, $\rho_0(k=0)$ gives the magnitude squared of the expectation value of the field.  In Fig.~\ref{fig:orderparameter}, we show the results for the temperature dependence of $\rho_0=\rho_0(k=0)$.  The superfluid density is proportional to $\rho_0$.  The proportionality constant, however, depends on the wave--function renormalization, which is not accurately determined so far.
\noindent
\parbox{\linewidth}{
\begin{figure}
\epsfig{file=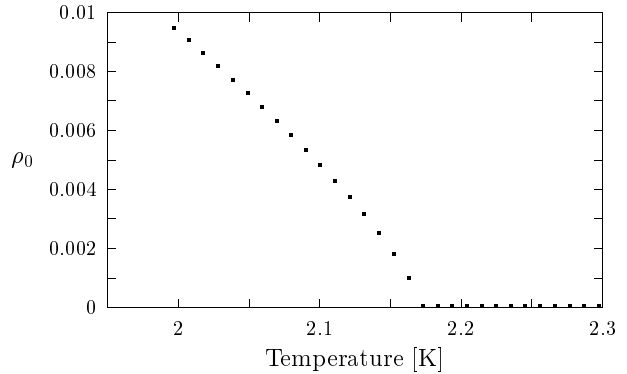, width = \linewidth}
\caption{Magnitude squared of the order parameter of the superfluid phase obtained from the evolution equation~(\ref{eq:dUprimedt}).  The chemical potential is adapted so that the calculated pressure yields the vapor pressure corresponding to the temperature.
\label{fig:orderparameter}}
\end{figure}
}
{
\bf Density.}
Finally, we compute the particle density as
\begin{eqnarray} \nonumber
n
&=&
\frac{1}{V} \Big< \sum_qa^\dag_qa_q \Big> \\ \label{eq:rawDensity}
&=&
\frac{1}{V}\sum_{n,q} \Big< \chi^*_{n,q}\chi_{n,q}\Big>
=
\frac{\rho_0}{V}+\frac{1}{2V}\mbox{Tr}\left(\Gamma^{(2)}_{k=0}\right)^{-1}.
\end{eqnarray}
The inverse full propagator is block diagonal in $n$ and $q$ with
\begin{equation}
\Gamma^{(2)}_{k}(n,q)
=
\left(
\begin{array}{cc}
\frac{\delta^{2}\Gamma_k[\varphi,\varphi^{*}]}{\delta\varphi^{*}_{n,q} \delta\varphi_{n,q}} &
\frac{\delta^{2}\Gamma_k[\varphi,\varphi^{*}]}{\delta\varphi_{-n,-q} \delta\varphi_{n,q}} \\
& \\
\frac{\delta^{2}\Gamma_k[\varphi,\varphi^{*}]}{\delta\varphi^{*}_{n,q} \delta\varphi^{*}_{-n,-q}} &
\frac{\delta^{2}\Gamma_k[\varphi,\varphi^{*}]}{\delta\varphi_{-n,-q} \delta\varphi^{*}_{-n,-q}} \\
\end{array}
\right)
\end{equation}
and contains for $k=0$ the information about the phonon energy spectrum for given $\mu$ and $T$ and arbitrary strength of the interaction.  The trace in Eq.~(\ref{eq:rawDensity}) includes a summation over momenta and Matsubara frequencies.  We calculate the eigenvalues of $\Gamma_{k=0}^{(2)}$ from the truncation (\ref{eq:truncation}) and explicitly sum over the Matsubara frequencies.  This yields (in the rescaled field variable)
\begin{equation}\label{eq:density}
n=\frac{2m}{\beta}\rho_0+\frac{1}{8\pi^2}\int_0^\Lambda dq\; q^2\frac{P_\pi+P_\sigma}{\sqrt{P_\pi P_\sigma}}\coth\left(\frac{\beta}{4m}\sqrt{P_\pi P_\sigma}\right),
\end{equation}
where $P_\pi$ and $P_\sigma$ again denote the inverse propagators of the Goldstone and the radial mode as in Eq.~(\ref{eq:props}), but with $R_{k=0}(n,q)=0$.
The masses $U_0^\prime$ and $U_0^\prime+2U_0^{\prime\prime}\rho$ are taken at $\rho_0$, so that in the spontaneously broken phase we have $U_0^\prime(\rho_0)=0$ and in the symmetric phase $\rho_0=0$ and $P_\pi = P_\sigma$.

As in the calculation of the pressure, the calculations of the density do not give reliable results for the absolute value of the density, which strongly depends on the UV cutoff and is expected to get a sizeable contribution from three--particle interactions.  We observe that the momentum integral in Eq.~(\ref{eq:density}) is dominated by high momenta $q^2\approx \Lambda^2$ and is therefore strongly affected by the choice of the cutoff.  Furthermore, this implies that the precise momentum dependence of the propagators $P_\pi$ and $P_\sigma$ could also have a substantial influence on the value of the density.  The temperature dependence of the density, on the other hand, is dominated by $q^2\approx 2mT$.  We thus allow for an overall shift of the density and only calculate its temperature dependence.  The results are shown in Fig.~\ref{fig:density}.  They reproduce the qualitative picture of the density close to the phase transition.  The peak at $T_c$ is clearly visible as well as the slightly negative thermal--expansion coefficient for $T<T_c$.
\noindent
\parbox{\linewidth}{
\begin{figure}
\epsfig{file=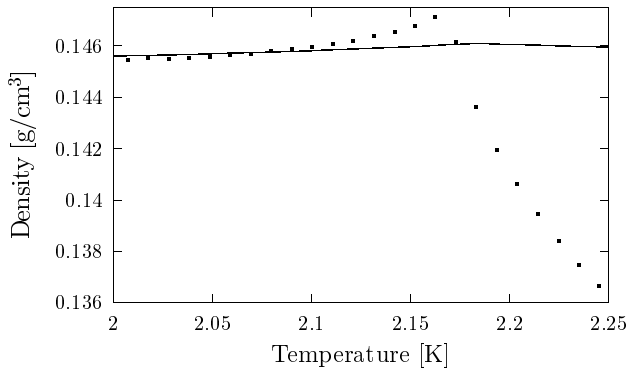, width = \linewidth}
\caption{Temperature dependence of the density.  The density is calculated for different temperatures around the critical point (filled boxes) and compared to measurements (solid line).  The chemical potential is adapted so that the calculated pressure yields the vapor pressure corresponding to the temperature.
\label{fig:density}}
\end{figure}
}

The fact that our calculations deviate substantially above the critical temperature is due to the crude treatment of the wave--function renormalization $Z$.  In the present approach, we have used $Z(q^2,T)=Z_\Lambda=1$ and $Y(q^2,T)=Y_\Lambda=0$.  For a more reliable treatment, the wave--function renormalizations $Z_{k=0}$ and $Y_{k=0}$ in $P_\pi$ and $P_\sigma$ should be replaced by momentum--dependent functions $Z(q^2,T)$, $Y(q^2,T)$.  We are aware that the simplifications $Z(q^2,T)=Z_\Lambda=1$ and $Y(q^2,T)=Y_\Lambda=0$ are less appropriate for the temperature dependence of the density, for which the integral is dominated by $q^2\approx 2mT$.    In the vicinity of the phase transition, the contribution to $dn/dT$ from the fluctuation integral (\ref{eq:density}) scales roughly $\propto Z^{-3/2}$.  Replacing $Z_\Lambda=1$ by $Z_{k_T}$, $k_T^2=2mT$ leads to a less pronounced temperature dependence if $Z_{k_T}>1$.  Furthermore, for $T$ near $T_c$, critical fluctuations influence the behavior of $Z$ and $Y$ at low $q^2$.  We believe that our crude treatment of the wave--function renormalization constitutes the most important source of error in determining the temperature dependence of the density.  We nevertheless see that we obtain qualitatively correct results, which are stable against shifting the UV cutoff.

Since the temperature dependence of the density is not a universal critical quantity, an understanding of the momentum and temperature dependence of $Z(q^2,T)$ is mandatory for a quantitative prediction.  This is illustrated by Fig.~\ref{fig:ZofT} where we show the temperature dependence of a momentum--independent $Z(T)$ that would be needed for a reproduction of the experimental data.  (The latter may be viewed as an effective $T$--dependent mean value of $Z(q^2,T)$, which yields the same $q^2$ integral~(\ref{eq:density}) as the true $Z(q^2,T)$.)  This curve seems reasonable, including the modest feature near $T_c$.  Using the dotted curve of Fig.~\ref{fig:ZofT} as an approximation to these $Z(T)$ values still leads to considerable deviations from the observed density in the immediate vicinity of $T_c$, demonstrating the high precision to which $Z$ has to be known.
\noindent
\parbox{\linewidth}{
\begin{figure}
\epsfig{file=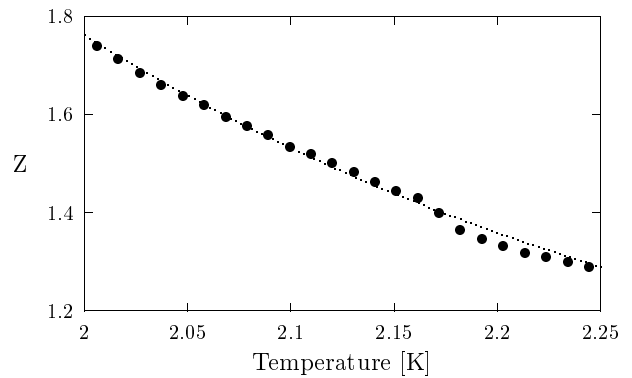, width = \linewidth}
\caption{Temperature dependence of the wave--function renormalization $Z$ (filled circles) that would be needed in Eq.~(\ref{eq:density}) to yield the observed density.  The dotted line displays a simple exponential fit to the data.
\label{fig:ZofT}}
\end{figure}
}

Another question why more precision on the wave--function renormalizations will be needed concerns the calculation of the excitation spectrum and the sound velocity.  This requires an understanding of the $q^2$ dependence of the inverse full propagator $\Gamma^{(2)}$ and thus strongly depends on $Z(q^2,T)$ and $Y(q^2,T)$.

\section{Conclusion}

We have computed thermodynamic properties of a strongly interacting system near a second--order phase transition from ``first principles''.  Starting from the microphysical interactions, we have determined the macroscopic thermodynamic potential for helium--4 near the superfluid phase transition.  Our method is based on an exact renormalization group equation.  The critical exponents obtained by this approach agree well with more elaborate state--of--the--art calculations and measurements.  Our most robust nonuniversal result, the critical chemical potential, is also in good agreement with experiment.  It supports our computation of the microphysical action by which helium--4 is described in the language of functional integrals.  This action contains an additive shift of the chemical potential that has not been considered so far.

For quantitatively convincing results of the equation of state, one needs to improve the treatment of the UV cutoff as well as additional effects of higher Matsubara frequencies beyond those that we have taken into account in the Schwinger--Dyson formalism.  A more accurate treatment of the wave--function renormalization is required as well.  Qualitatively, though, we reproduce important features of the equation of state such as the density peak at the phase transition.

We emphasize that we treat here a temperature range where collective effects and critical behavior are crucial and where many previous methods are plagued by severe infrared problems.  Despite certain quantitative shortcomings, this work may be viewed as a demonstration that \emph{ab initio} computations of the equation of state become possible with modern renormalization--group techniques.


\end{multicols}

\end{document}